\documentclass[twocolumn,showpacs,preprintnumbers,nofootinbib,prd,superscriptaddress,groupedaddress,10pt]{revtex4-1}

\usepackage{graphicx,amssymb,amsmath,amsthm,amsfonts,epsfig,epsf}
\usepackage[linktocpage]{hyperref}
\usepackage[usenames]{color}
\usepackage{epstopdf}

\usepackage{aas_macros}
\usepackage{bm}
\usepackage{dcolumn}
\usepackage[latin1]{inputenc}
\usepackage{latexsym}
\usepackage{rotating}
\usepackage{hyperref}
\usepackage{color}
\usepackage{longtable}
\usepackage{enumerate}
\usepackage{tensor}
\usepackage{mathtools}
\usepackage{url}
\newcommand{\ben}{\begin{enumerate}}
\newcommand{\een}{\end{enumerate}}

\def\be{\begin{equation}}
\def\ee{\end{equation}}
\def\bea{\begin{eqnarray}}
\def\eea{\end{eqnarray}}
\newcommand{\beq}{\begin{eqnarray}}
\newcommand{\eeq}{\end{eqnarray}} 
\newcommand{\ba}{\begin{align}}
\newcommand{\ea}{\end{align}}

\begin{document}

\title{Superradiance in stars}

\author{
Vitor Cardoso$^{1,2}$ 
,
Richard Brito$^{1}$
,
Jo\~ao L. Rosa$^{1}$
,
}
\affiliation{${^1}$ CENTRA, Departamento de F\'{\i}sica, Instituto Superior T\'ecnico -- IST, Universidade de Lisboa -- UL,
Avenida Rovisco Pais 1, 1049 Lisboa, Portugal}
\affiliation{${^2}$ Perimeter Institute for Theoretical Physics Waterloo, Ontario N2J 2W9, Canada}


\begin{abstract}
It has long been known that dissipation is a crucial ingredient in
the superradiant amplification of wavepackets off rotating objects.
We show that, once appropriate dissipation mechanisms are included, stars are also prone to superradiance and superradiant instabilities.
In particular, ultra-light dark matter with small interaction cross
section with the star material or self-annihilation can trigger a superradiant instability.
On long timescales, the instability strips the star of most of its angular momentum.
Whether or not new stationary configurations surrounded by scalar condensates exist, remains to be seen.
\end{abstract}



\maketitle

\section{Introduction}

Rotating bodies with internal degrees of freedom -- where energy can be dumped into -- display superradiance.
This means that the scattering of low-frequency waves extracts rotational energy away from the body and is used to amplify incoming radiation~\cite{zeldovich2,Bekenstein:1998nt,Brito:2015oca}. When the object is a black hole, dissipation is intrinsically built in the problem under the form of a one-way membrane: the event horizon. Black hole superradiance was in fact shown to exist and to trigger interesting phenomena.
These include black hole bombs, once the black hole is surrounded by some confining medium, giving rise to exponentially growing modes~\cite{Press:1972zz,Cardoso:2004nk,Cardoso:2004hs,Cardoso:2006wa,Cardoso:2013pza}. The scattering of massive fields belongs to this category:
the mass term effectively confines the field giving rise to floating orbits~\cite{Cardoso:2011xi} or to superradiant instabilities, which
extract rotational energy away from the black hole~\cite{Damour:1976kh,Detweiler:1980uk,Cardoso:2005vk,Pani:2012vp,Witek:2012tr,Brito:2013wya,Cardoso:2013fwa}.
The instability leads to a depletion of mass and angular momentum from the black hole, resulting in observationally-interesting ``holes'' in the Regge-plane~\cite{Arvanitaki:2010sy,Brito:2014wla,Arvanitaki:2014wva}. As a consequence, the observation of supermassive black holes alone can impose stringent bounds on the mass of ultralight bosons~\cite{Pani:2012vp,Brito:2013wya}. Finally, in multi-scalar theories or theories with a complex scalar, superradiance is the responsible for new hairy black hole solutions~\cite{Hod:2012px,Herdeiro:2014goa,Herdeiro:2015gia} (whether or not these solutions are generically approached as the end-state of the superradiant-instability is discussed in Ref.~\cite{Brito:2014wla}).
For a Review, we refer the reader to Ref.~\cite{Brito:2015oca}.

Given the rich phenomenology of black hole superradiance, one might ask whether those results carry through to other self-gravitating objects.
Ideal stars do not display superradiance and none of the associated instabilities (except for possible ergoregions instabilities~\cite{Brito:2015oca}). But realistic stars {\it do} dissipate energy and they have several channels available for it. 
The original arguments by Zeldovich already made clear that any classical system with dissipation will be prone to superradiant scattering~\cite{zeldo1,zeldovich2} (see also Ref.~\cite{Glampedakis:2013jya} who studied the correspondence between superradiance and tidal friction on viscous Newtonian anisotropic stars). In fact, stars in full General Relativity were recently shown to display superradiance, as long
as dissipation is included~\cite{Richartz:2013unq}. The purpose of this work is to compute explicitly the amplification factors when a toy model for dissipation is used, and to show that, within this toy model, stars are superradiantly-unstable against massive scalar-field fluctuations.

Our results make {\it possible} the existence of ``hairy'' stars~\footnote{by which we mean new stationary configurations describing
a star surrounded by a scalar-field condensate}, supported solely by superradiance.
For dark matter models comprised of light degrees of freedom, our results open the door to constrain the interaction cross section
with ordinary matter or the self-annihilation cross-section.

\section{Setup}
We are interested in describing ultra-light bosonic degrees of freedom, described by
a scalar minimally coupled to gravity, and interacting with a star. Dissipation can be modelled in several -- and complex -- ways.
For instance, one can let the scalar couple to spacetime degrees of freedom, or one can consider
the coupling to microscopic degrees of freedom, describing the interaction of self-annihilation of -- for example -- dark matter with a star.
The description of a dissipative system within full General Relativity for systems (such as stars) with many internal degrees of freedom
is a very complex problem~\cite{Andersson:2013jga}.

Instead, we follow Zeldovich's seminal work and consider a toy model for absorption, by modifying
the Klein-Gordon equation inside the star {\it and in a frame co-rotating with the star} as~\cite{zeldo1}
\be
\Box \Phi+\alpha\frac{\partial \Phi}{\partial t}=\mu^2\Phi\,. \label{KGdiss}
\ee
Here, ${\mu}\hbar$ is the mass of the scalar field.
The $\alpha$ term is added to break Lorentz invariance, and describes absorption on a timescale $\tau \sim 1/\alpha$,
in a frame co-rotating with the star. 
The constant $1/\alpha$ has dimensions of time, or length in geometrical units.
Contact with particle physics can be made by considering the particle associated with the scalar field and identifying $1/\alpha$ with the mean free path $\ell$ of the particle inside the star. In other words, for a particle with cross section $\sigma$ to decay into other particles via interaction with say a neutron of mass $m_N$, then
\be
\alpha=n\sigma\,,
\ee
where $n$ is the nucleon number density in the star. One can also express the dissipation parameter $\alpha$ with the help of the star's total mass $M$, in the dimensionless combination
\beq
M\alpha&=&\frac{GM}{c^2}\frac{\rho \sigma}{m_N}\nonumber\\
&\sim& 0.92 \,\frac{\rho}{10^{15}\,{\rm Kg\,m}^{-3}}\frac{M}{M_{\odot}}\frac{\sigma}{10^{-41}\,{\rm cm}^2}\,.
\eeq
Thus, the dimensionless combination can be of order one for interesting dark matter candidates~\cite{Gould:1987ir,Gould:1987ju,Macedo:2013qea} (and references therein).
The constant $\alpha$ could also be used similarly to describe dark matter self-annihilation.

It is easy to see, following Zeldovich, that if the frequency in the accelerated frame is $\omega$ and the field behaves as $e^{-i\omega t+im\varphi}$ then in the inertial frame the azimuthal coordinate is $\varphi=\varphi'-\Omega t$, and hence the frequency is $\omega'=\omega-m\Omega$. In other words, the effective damping parameter $\alpha \omega'$ becomes negative in the superradiant regime and the medium amplifies -- rather than absorbing-- radiation~\cite{zeldo1,zeldovich2,Brito:2015oca}.
This procedure can be used to derive the wave equation of a scalar field in a rotating black hole background (up to first order in rotation), starting from that of a scalar field in a Schwarzschild geometry.
As we will shortly see, it also allows to predict the amplification factors of scalar fields in a Kerr background. Thus, while only a phenomenological approach to the problem,
the inclusion of the Lorentz-violating term is well motivated and consistent with known results. In Appendix~\ref{sec:liberati} we work out another model for dissipation, inspired in the theory behind dissipation in fluids~\cite{Liberati:2013usa}. The results are perfectly consistent with the toy model~\eqref{KGdiss} adopted in the main body of this work.

As is clear from the previous discussion, superradiance requires only rotation and dissipation. The particular background geometry or material
is only relevant to quantitatively describe the phenomena. We will focus exclusively on backgrounds describing a constant density, perfect fluid star in General Relativity as a proxy for the generic case:
\be
ds^2=-e^{2\Phi}dt^2+\frac{dr^2}{1-2m(r)/r}+r^2d\Omega^2\,,
\ee
with 
\beq
m&=&\frac{4\pi}{3}\rho r^3\,,\,\,\,P=\rho\left(\frac{\sqrt{1-2Mr^2/R^3}-\sqrt{1-2M/R}}{3\sqrt{1-2M/R}-\sqrt{1-2Mr^2/R^3}}\right)\,,\nonumber\\
e^{\Phi}&=&\frac{3}{2}\sqrt{1-2M/R}-\frac{1}{2}\sqrt{1-2Mr^2/R^3}\,.
\eeq
Here, $M,R$ are the star's mass and radius, and are related to the density $\rho$ via $M=\frac{4\pi}{3}\rho\,R^3$.

Notice that we take an exact general-relativistic solution describing a {\it non-rotating} star. Although not self-consistent, the effects of General Relativity
are subdominant. The leading effect of rotation is taken into account when transforming to a frame at rest with the star, and using \eqref{KGdiss}.

\section{Superradiant amplification factors of massless fields}
\begin{figure}[ht]
\begin{center}
\epsfig{file=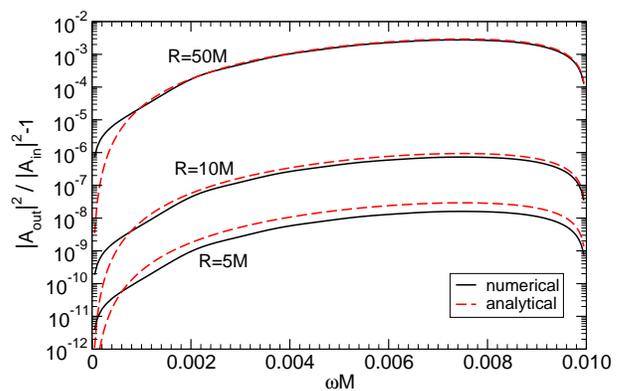,width=8cm,angle=0,clip=true}
\end{center}
\caption{\label{fig:superradiance} Superradiant amplification of a scalar field by a rotating star, where dissipation is modeled through equation \eqref{KGdiss}.
Here, $M\alpha=0.1,\,M\Omega=0.01$.
We find that the amplification factor scales with the dissipation parameter $\alpha$ for small $M\alpha$. For non-compact, slowly-rotating stars, the amplification factor (solid lines) is well described by the analytic result~\eqref{analytic},
shown as a dashed curve in the plot. 
}
\end{figure}
To solve the Klein-Gordon equation \eqref{KGdiss}, we use the following ansatz for the field,
\be \label{ansatz}
\Phi(t,r,\vartheta,\varphi)=\frac{\Psi(r)}{r}e^{-i\omega t+im\varphi}P_l(\cos\vartheta)\,,
\ee
where $P_l$ are Legendre functions. To solve the equation inside the star, we first transform to a frame co-rotating with the star (which amounts to replacing $\varphi\to \varphi-\Omega t$, or $\omega\to \omega-m\Omega$),
impose regularity at the center and then match the radial function and its derivative. For now we focus on zero field mass, $\mu=0$. At large distances, the solution takes the form
\be
\Psi=A_{\rm out}e^{i\omega r}+A_{\rm in}e^{-i\omega r}\,,
\ee
where the subscripts denote an out-going and an in-going piece (due to the time dependence of the ansatz for the field).
Thus, the asymptotic solution allows one to frame this in a context of a scattering experiment. We define the amplification factor as
$|A_{\rm out}|^2/|A_{\rm in}|^2-1$. Positive amplification factors correspond to superradiant amplification from the star.
Negative factors correspond to overall absorption by the star.

Our numerical results are summarized in Fig.~\ref{fig:superradiance} for $M\alpha=0.1,\,M\Omega=0.01$. Superradiant amplification does exist, as was shown generically in Ref.~\cite{Richartz:2013unq},
and the amplification can be significant. In fact, we find that it scales linearly with $M\alpha$ for small $M\alpha$ and it increases significantly as the star's surface velocity increases.
Amplification factors can be of order of those around rotating Kerr black holes or higher.

For non-relativistic, Newtonian configurations ($M/R,\,\Omega R\ll 1$) the wave equation can be solved analytically inside and outside the star
in terms of Bessel functions. We find a simple analytic expression for the amplification factor,
\be
\frac{|A_{\rm out}|^2}{|A_{\rm in}|^2}-1=\frac{4\alpha R\left(R\Omega-\omega R\right)\left(\omega R\right)^{2l+1}}{(2l+1)!!(2l+3)!!}\,.\label{analytic}
\ee
As can be seen from Fig.~\ref{fig:superradiance}, this relation gives a very good approximation to the numerical results for a good fraction of the parameter space, including relatively compact stars.
This relation is also interesting, as it allows one to predict the amplification factor for rotating {\it black holes}. For black holes, $R\sim 2M$ and $1/\alpha=M$ is the only possible timescale in the problem. For example, the above relation predicts that slowly rotating black holes in General Relativity amplify $l=1$ scalar fields with
\be
\frac{|A_{\rm out}|^2}{|A_{\rm in}|^2}-1=\frac{16}{45}M\left(\Omega-\omega\right)\left(2M\omega \right)^{3}\,.\label{analytic2}
\ee
On the other hand, a matched-asymptotic expansion calculation in full General Relativity yields approximately the same result (the coefficient turns out to be $2/9\sim 0.22$ instead of $16/45\sim0.355$~\cite{1973ZhETF..65....3S,Brito:2015oca}). 

\section{Superradiant instability against massive fields}
As soon as the field is allowed to be massive, superradiance can trigger new phenomena, in particular instabilities:
the mass confines the field within a distance $1/\mu^2$ of the star, whereas superradiance amplifies it. This effect was shown to occur for black holes~\cite{Damour:1976kh,Detweiler:1980uk,Cardoso:2005vk,Pani:2012vp,Witek:2012tr,Brito:2013wya,Cardoso:2013fwa,Brito:2015oca},
we now show that it occurs generically.

For massive fields, the asymptotic behavior at spatial infinity is of the form $e^{\pm \sqrt{\mu^2-\omega^2}\,r}$.
Thus, it is possible that some eigenvalues $\omega=\omega_R+i\omega_I$ exist for which the field is regular at infinity (for $\omega_R<\mu$) and for which the mode
is unstable (for $\omega_I>0$, cf. Eq. \eqref{ansatz}). We have used a direct integration approach to search for these modes.

\begin{figure*}[htb!]
\begin{center}
\begin{tabular}{cc}
\epsfig{file=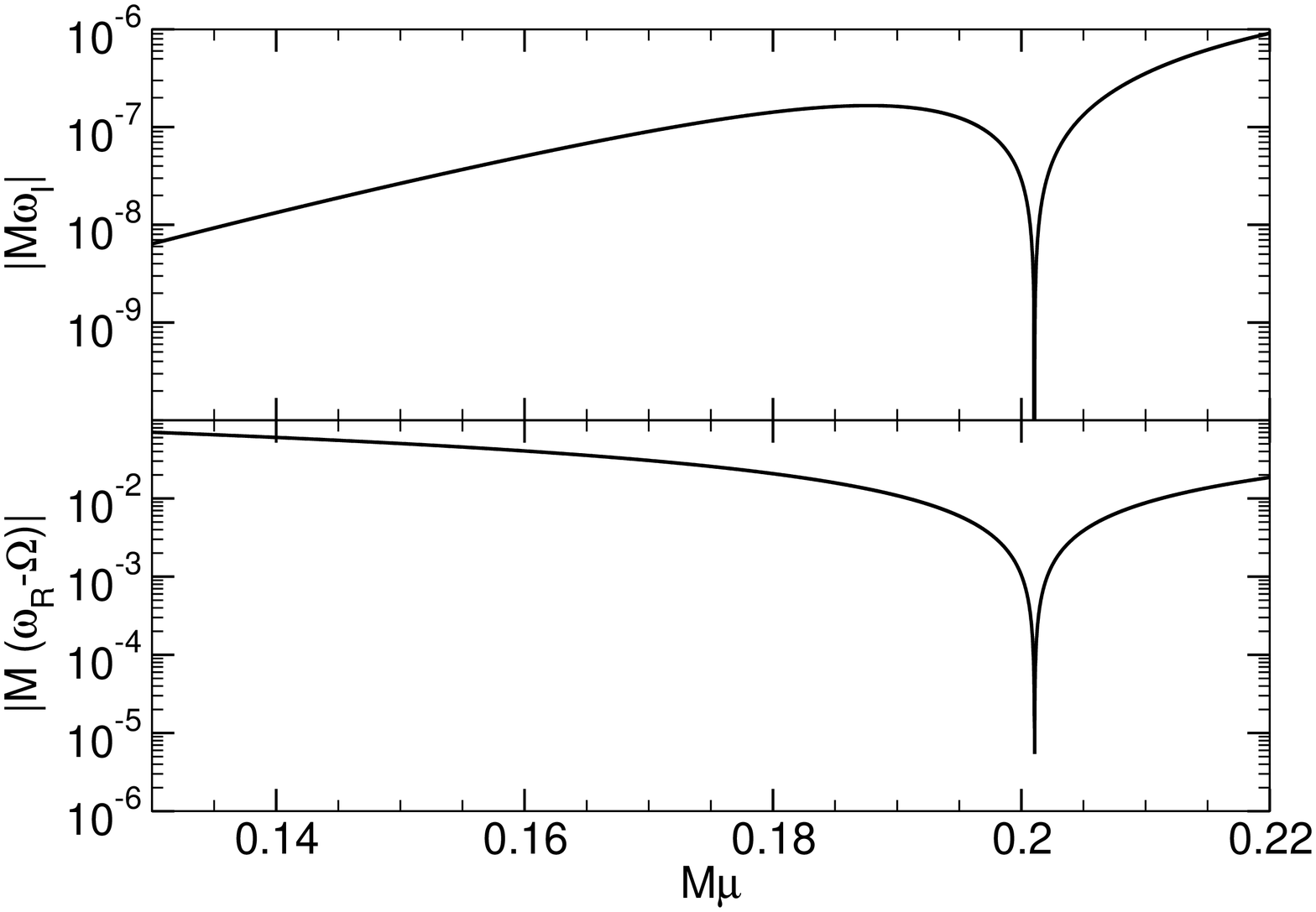,width=8cm,angle=0,clip=true}&\epsfig{file=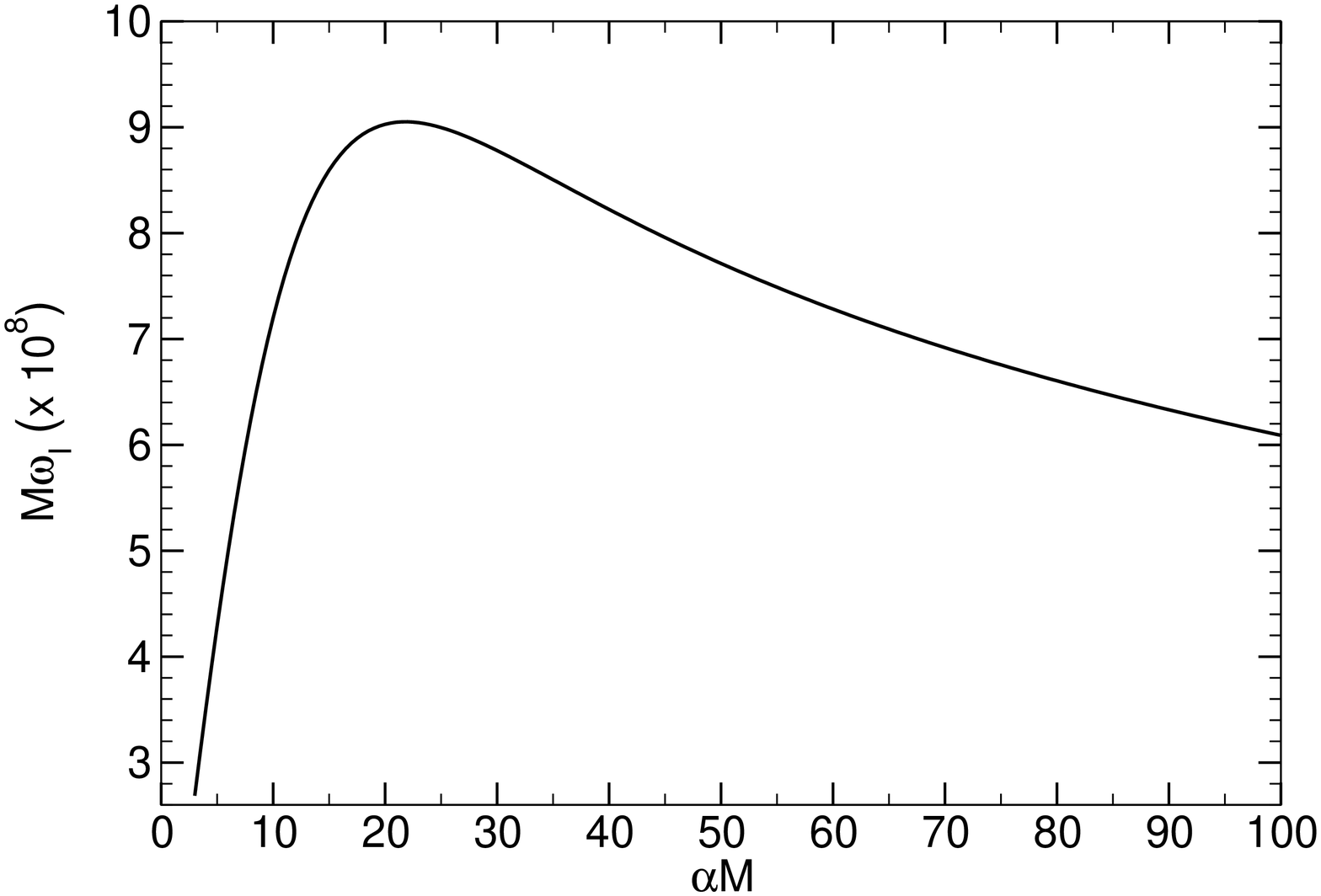,width=8cm,angle=0,clip=true}
\end{tabular}
\end{center}
\caption{\label{fig:instability} Superradiant instability of a massive scalar field coupled to a rotating star, where dissipation is modeled through equation \eqref{KGdiss}.
Here, $R/M=4,\,M\Omega=0.2$, $M\alpha=20$ (left panel) and $M\mu=0.17$ (right panel). The transition from instability $\omega_I>0$ to stability $\omega_I<0$ coincides with a change in sign of $\omega_R-m\Omega$. We find that $M\omega_I$ scales linearly with the dissipation parameter $\alpha$ for small $M\alpha$.
}
\end{figure*}

Figure~\ref{fig:instability} summarizes our results for the instability of massive scalars coupled to dissipative stars.
For a fixed star radius and mass $(R,\,M)$ and dissipation parameter $\alpha$, there {\it are} unstable modes. At very small values of $M\mu$, the instability timescale
decreases when the scalar mass $\mu$ increases. We find that at low masses $\mu$ our results are well described by $\omega_I\propto (\omega_R-m\Omega)(M\mu)^{\eta}$ with $\eta \sim 9\pm 2$.
However, the real part of the mode $\omega_R\sim \mu$ also increases. Thus, for $\mu\sim \Omega$ the $m=1$ mode ceases to be unstable. This is clearly seen in Fig.~\ref{fig:instability},
which shows that the transition from instability to stability coincides with a change in sign of $\omega_R-m\Omega$.

%

\section{Discussion}
Neutron stars or other highly compact stars with dissipative channels may be interesting probes
of physics beyond the Standard Model, specially of bosonic fields with a Compton wavelength of the order of the star size.
Our working model is extremely crude and more sophisticated ways of including dissipation should be sought for. Nevertheless,
our work certainly shows that there may be a payoff to these efforts: 
highly spinning compact stars develop instabilities against massive fields
which may be used (see e.g. Refs.~\cite{Brito:2014wla,Brito:2015oca}) to further constrain ultralight bosonic fields, while the bosonic clouds formed during the process might emit very distinctive gravitational-wave signals within the optimal sensitivity band of ground-based detectors~\cite{Arvanitaki:2014wva}. It is even possible, but it remains to be proven in a concrete theory with dissipation, that new star-like configurations exist, which would be the counterpart to hairy black hole solutions recently studied in theories with a minimally coupled massive scalar~\cite{Hod:2012px,Herdeiro:2014goa,Herdeiro:2015gia,Berti:2015itd}.
\begin{acknowledgments}
V.C. acknowledges financial support provided under the European
Union's FP7 ERC Starting Grant ``The dynamics of black holes: testing
the limits of Einstein's theory'' grant agreement no. DyBHo--256667,
and FP7 ERC Consolidator Grant ``Matter and strong-field gravity: New frontiers in Einstein's theory'' grant agreement no. MaGRaTh--646597.
R.B. acknowledges financial support from the FCT-IDPASC program
through the Grant No. SFRH/BD/52047/2012, and from the Funda\c c\~ao
Calouste Gulbenkian through the Programa Gulbenkian de Est\' imulo \`a
Investiga\c c\~ao Cient\'ifica.
This research was supported in part by the Perimeter Institute for
Theoretical Physics. Research at Perimeter Institute is supported by
the Government of Canada through Industry Canada and by the Province
of Ontario through the Ministry of Economic Development $\&$
Innovation.
This work was supported by the NRHEP 295189 FP7-PEOPLE-2011-IRSES
Grant.
\end{acknowledgments}
\vskip 5cm
\appendix
\section{\label{sec:liberati}Superradiant amplification factors in a viscous gravity analogue}
%
\begin{figure}[ht]
\begin{tabular}{c}
\epsfig{file=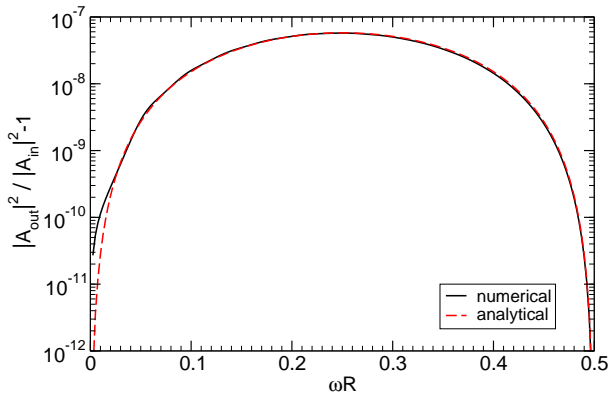,width=8cm,angle=0,clip=true}
\end{tabular}
\caption{\label{fig:liberati} Superradiant amplification of a scalar field by a rotating star, where dissipation is modeled through equation \eqref{fluid0}.
Here, $\frac{4}{3R}\nu=0.002,\,R\Omega=0.5$.
}
\end{figure}
A different way to model dissipation in a gravitational system is to use the methods of analogue gravity. For example, linear perturbations in an inviscid, irrotational flow propagate as fields on a curved space-time. Thus, one natural option to model dissipative phenomena is through the inclusion of a kinematic viscosity. For a fluid at rest, the wave equation for the perturbations is given by~\cite{Liberati:2013usa}
\be 
\label{fluid0}
\partial_t^2\phi=\nabla^2\phi+\frac{4}{3}\nu\partial_t\nabla^2\phi\,.
\ee
%
%
%
Thus, we will here briefly describe how our results change when the dissipation inside the star is modeled through \eqref{fluid0}.
We are here concerned only with rotational superradiance in flat spacetime, and we therefore take $\nabla^2$ to be the flat-space Laplacian in spherical coordinates. 
%
The process of transforming to a co-rotating frame and the imposition of regularity at the center are identical to that of the model studied in the main text. 
The amplification factors were numerically obtained and are shown in Fig. \ref{fig:liberati}. 
Note that for low frequencies it is possible to map this analogue model into the previous toy model. Equation \eqref{fluid0} can be written as
\be
\left(1-\frac{4}{3}i\omega\nu\right)\Box\phi+\frac{4}{3}i\omega\nu\partial_t^2\phi=0,
\ee
which in the limit $\omega\ll 1$ is the same as Eq.~\eqref{KGdiss}, for $\alpha=\frac{4}{3}\nu\omega^2$. This map between $\alpha$ and $\nu$ allows us to obtain the non-relativistic analytical expression for the amplification factors from equation \eqref{analytic}:
\be
\frac{|A_{\rm out}|^2}{|A_{\rm in}|^2}-1=\frac{16\nu R^2\left(\Omega-\omega\right)^3\left(\omega R\right)^{2l+1}}{3(2l+1)!!(2l+3)!!}\,.\label{analyticlib}
\ee 

It is worth noting that, if one were to try the naive black hole limit $R=2M$ with $\nu$ of order $M$, the above result would never recover the correct behavior for the amplification
of waves by rotating black holes. The reason, it seems, is that dissipation in non-rotating black holes can be best described through \eqref{KGdiss} rather than this viscous analogue. This is possibly due to the fact that the dispersion relation of Eq.~\eqref{fluid0} in the eikonal limit leads to an energy-dependent dissipative term~\cite{Liberati:2013usa}.

\bibliographystyle{h-physrev4}
\bibliography{Star_extraction_refs}

\end{document}